\documentclass[%
 reprint,
 amsmath,amssymb,
 aps,
superscriptaddress
]{revtex4-2}

\usepackage{graphicx}
\usepackage{dcolumn}
\usepackage{bm}
\usepackage{diagbox}

\begin{document}

\title{Architectural Foundations for Checkpointing and Restoration in Quantum HPC Systems}%

\author{Qiang Guan}
\affiliation{Kent State University, Kent, Ohio, United States}
\email{qguan@kent.edu}

\author{Qinglei Cao}
\affiliation{Saint Louis University, St. Louis, Missouri, United States}
\email{qinglei.cao@slu.edu}

\author{Xiaoyi Lu}
\affiliation{University of Florida, Gainesville, Florida, United States}
\email{xiaoyilu@ufl.edu}

\author{Siyuan Niu}
\affiliation{University of Central Florida, Orlando, Florida, United States}
\email{siyuan.niu@ucf.edu}

\date{\today}

\begin{abstract}

In this work, we explore the design of the checkpointing and restoration for quantum HPC that leverages dynamic circuit technology to enable restartable and resilient quantum execution. Rather than attempting to checkpoint quantum states, our approach redefines checkpointing as a control flow and algorithmic state problem. By exploiting mid-circuit measurements, classical feed forward, and conditional execution supported by dynamic circuits, we capture sufficient program state to allow correct restoration of quantum workflows after interruption or failure. This design aligns naturally with iterative and staged quantum algorithms such as variational eigensolvers, quantum approximate optimization, and time-stepping methods commonly used in quantum simulation and scientific computing.
\end{abstract}


\maketitle

\section{Introduction}
Quantum computing is increasingly integrated into HPC environments to support hybrid workflows ~\cite{Yuqi-SC25,Yuqi-advancedscience} in chemistry, materials science, optimization, and scientific simulation. These applications often involve long-running iterative loops, large classical preprocessing steps, and repeated quantum executions. In classical HPC, resilience is achieved through checkpointing and rollback mechanisms that preserve application state across failures. However, direct analogues of these mechanisms do not exist for quantum programs, as quantum states cannot be duplicated or restored once lost. 

Long-running quantum applications executed in hybrid quantum classical high performance computing environments face fundamental challenges in reliability and fault recovery. Unlike classical HPC systems where application state can be checkpointed through memory snapshots, quantum states cannot be copied or preserved due to the no-cloning theorem and measurement-induced collapse. As a result, existing checkpointing techniques are fundamentally incompatible with quantum execution, limiting the scalability and robustness of quantum HPC workflows. The differences in checkpointing features in HPC and Quantum-HPC has been shown in table ~\ref{tab:diff}.

\begin{table}[t]
\centering
\small
\resizebox{\columnwidth}{!}{%

\begin{tabular}{|c|c|c|c|c|c|c|}
\hline
\diagbox[width=3cm]{\textbf{Checkpoints}}{\textbf{Features}} & \textbf{\begin{tabular}{@{}c@{}}State\\Clonability\end{tabular}} & \textbf{\begin{tabular}{@{}c@{}}Full State\\Sorability\end{tabular}} & \textbf{\begin{tabular}{@{}c@{}}Full State\\Restoration\end{tabular}} & \textbf{\begin{tabular}{@{}c@{}}Time\\Constraint\end{tabular}} & \textbf{\begin{tabular}{@{}c@{}}HW Metadata\\Storage\end{tabular}} & \textbf{\begin{tabular}{@{}c@{}}MCM\\Support\end{tabular}} \\
\hline
HPC & $\checkmark$ & $\checkmark$ & $\checkmark$ & $X$ & $X$ & $X$\\
\hline
Quantum & $X$ & $X$ & $X$ & $\checkmark$ & $\checkmark$ & $\checkmark$ \\
\hline
\end{tabular}
}
\caption{Comparison of different checkpoint features in HPC and quantum settings.}
\vspace{-6mm}
\label{tab:diff}
\end{table}

Recent advances in dynamic quantum circuits~\cite{corcoles2021exploiting} introduce a critical opportunity to revisit this limitation. Dynamic circuits enable mid-circuit measurements, conditional branching, and real-time classical feedback within a single quantum program. These capabilities effectively introduce a notion of program state and control flow into quantum execution, allowing quantum algorithms to adapt their behavior based on intermediate outcomes. This work explores how dynamic circuits can be systematically exploited to enable checkpointing and restoration semantics suitable for quantum HPC.

We introduce a checkpointing framework built on three core principles: (i) checkpointing is defined at the level of algorithmic and control-flow state rather than quantum state, (ii) dynamic circuits are used to convert selected quantum information into classical representations through structured measurements at well-defined program boundaries, and (iii) restoration is achieved through controlled re-execution of quantum circuits guided by recorded classical state and parameters. The framework supports multiple classes of checkpoints, including classicalized checkpoints that capture measurement outcomes and algorithmic metadata such as iteration counters and parameter values, algorithmic checkpoints that align with natural phase boundaries in iterative quantum algorithms to enable safe restart from completed stages, and extensions to logical checkpoints in fault-tolerant settings that incorporate error syndrome histories and decoder state to support logical-level restoration. A quantum HPC runtime coordinates checkpoint creation, storage, and restoration in concert with classical schedulers and failure detectors, and upon failure or preemption, reconstructs execution by re-instantiating circuits, rehydrating parameters, and conditionally replaying execution paths using dynamic circuit control.


\vspace{-3mm}\section{Architectural Design}
This work proposes a layered architecture that enables checkpointing and restoration for quantum high performance computing workflows by leveraging dynamic circuit capabilities. Rather than attempting to preserve quantum states, which is fundamentally prohibited, the architecture redefines checkpointing as the capture and restoration of algorithmic and control flow state across quantum and classical execution layers. The architecture is designed to integrate seamlessly with existing HPC runtime systems while respecting quantum mechanical constraints.

\noindent\textbf{Quantum HPC Runtime and Control Layer}. At the core of the architecture is a quantum HPC runtime and control layer that orchestrates checkpoint creation, failure detection, and restoration. This layer functions analogously to a classical HPC runtime system but is explicitly aware of quantum execution semantics and dynamic circuit capabilities. The runtime includes a checkpoint manager responsible for identifying safe checkpoint boundaries based on algorithmic structure, execution progress, and system policies. Checkpoints may be triggered at iteration boundaries, circuit layer boundaries, convergence points, or runtime events such as preemption or failure. A restoration engine reconstructs quantum execution after interruption by re-instantiating quantum circuits, rehydrating parameters and metadata, and conditionally replaying execution paths using dynamic circuit control. A failure detection and policy engine continuously monitors quantum backend availability and classical system health, enabling adaptive responses such as rollback, restart, or rescheduling. 

\noindent\textbf{Quantum Program Layer with Dynamic Circuits}. The quantum program layer provides the execution substrate that makes checkpointing feasible through dynamic circuits supporting mid-circuit measurement, conditional branching, and classical feedforward. These features introduce program state and control flow into quantum execution, structuring programs into execution regions separated by checkpoint boundaries. Within each region, quantum operations proceed until a checkpoint point is reached, where measurements project selected quantum information into classical form and export it to the classical runtime. Subsequent execution depends on recorded measurement outcomes through conditional operations, ensuring algorithmically consistent restoration even though the underlying quantum state is not preserved.
This structure can accommodate multiple checkpoint classes through different dynamic circuit patterns. For qubit reuse, classicalized checkpoints store final measurement outcomes before qubits are reset and reused, enabling restart on reduced qubit layouts~\cite{hua2023caqr,decross2023qubit,niu2024effective}. In dynamic state preparation (e.g., constant-depth GHZ, W-states)~\cite{baumer2024efficient,farrell2025digital, niu2024ac}, checkpoints align with probabilistic measurement branches, storing outcomes to reconstruct the preparation path. For variational algorithms like Feedback-based Algorithm for Quantum Optimization (FALQON)~\cite{magann2022feedback}, algorithmic checkpoints capture measurement results, preserving adaptive ansatz construction. In fault-tolerant settings, logical checkpoints store syndrome measurements and decoder states, enabling the restoration of error tracking and decoding status. 

\noindent\textbf{Classical HPC Layer and Checkpoint Storage}. All checkpoint data is stored and managed within the classical HPC layer, which leverages existing checkpoint storage infrastructure~\cite{scr-moody,raghu-hpdc13,nvme-cr}. Stored data includes measurement outcomes, variational parameters, iteration counters, control flow decisions, random seeds, and hardware calibration metadata. From the perspective of the HPC system, these checkpoints are structured metadata objects rather than memory snapshots, allowing them to be stored using conventional parallel file systems or burst buffers.
This design preserves compatibility with existing HPC schedulers and resource managers, enabling quantum workloads to participate in standard resilience and preemption mechanisms without requiring specialized quantum state storage.
\vspace{-3mm}\section{Feasibility}
The proposed checkpointing and restoration architecture is feasible in the near term due to recent advances in dynamic quantum circuit execution and the maturity of classical HPC runtime infrastructure. Dynamic circuit capabilities such as mid circuit measurement, conditional branching, and classical feed forward are already supported on emerging quantum hardware platforms and are being exposed through compiler stacks and intermediate representations, including QIR, MLIR, and CUDA Q. These capabilities provide the minimal primitives required to encode program state and control flow without violating quantum mechanical constraints.

From a systems perspective, the architecture does not require new quantum hardware features beyond those already available or on near-term roadmaps. Instead, it leverages existing measurement operations and classical control paths in a structured manner. All checkpoint data are classical in nature and are stored using standard HPC checkpoint mechanisms, making the approach compatible with existing schedulers, burst buffers, and parallel file systems. As a result, the integration effort is concentrated at the runtime and compiler levels rather than at the hardware layer, significantly reducing implementation risk.

The framework is particularly well-suited to iterative and staged quantum algorithms such as variational eigensolvers, quantum approximate optimization, time stepping methods, and hybrid quantum classical workflows. These algorithms naturally expose safe checkpoint boundaries at iteration or layer transitions, allowing checkpointing to be introduced with minimal algorithmic restructuring. While checkpointing introduces additional measurement overhead and partial loss of coherence, this overhead is predictable, controllable, and amortized over long-running executions, making it acceptable for workloads where resilience and restartability are critical.
Finally, the design admits incremental deployment. Initial prototypes can focus on classicalized and algorithmic checkpoints for near-term hardware, while the same architectural abstractions can later be extended to logical qubit checkpointing in fault-tolerant regimes. This staged feasibility path enables meaningful validation and performance characterization at each stage of hardware evolution.
\vspace{-3mm}\section{Conclusion}
In this work, we redefine checkpointing for quantum HPC systems by focusing on capturing and restoring algorithmic and control-flow state using dynamic quantum circuits rather than preserving quantum states. Leveraging mid-circuit measurements, classical feed-forward, and conditional execution, the proposed architecture enables restartable and resilient quantum workflows while remaining compatible with quantum mechanical constraints and existing HPC runtimes. The approach naturally supports iterative and staged quantum algorithms, where checkpoint boundaries align with algorithmic phases and overheads are predictable, and provides a practical near-term solution with a clear path toward logical-level checkpointing in future fault-tolerant systems.

\bibliographystyle{unsrt}
\bibliography{siyuan,guan,luxi}
\end{document}